\documentclass[journal,transmag]{IEEEtran}
\usepackage{color}
\usepackage{graphicx}
\usepackage{amsmath}
\usepackage{subcaption}
\usepackage{upgreek}
\usepackage{units}
\usepackage[colorinlistoftodos,prependcaption,textsize=tiny]{todonotes}
\usepackage{amssymb} %for use of \mathbb{N} for Element-symbols

\usepackage{epstopdf}
\ifCLASSINFOpdf
\else
\fi
\begin{document}
\title{Spin-current manipulation of photoinduced magnetization dynamics in heavy metal / ferromagnet double layer based nanostructures}


\author{\IEEEauthorblockN{Steffen Wittrock\IEEEauthorrefmark{1},
Dennis Meyer\IEEEauthorrefmark{2},
Markus M\"{u}ller\IEEEauthorrefmark{2}, 
Henning Ulrichs\IEEEauthorrefmark{2},
Jakob Walowski\IEEEauthorrefmark{3},
Maria Mansurova\IEEEauthorrefmark{3},\\
Ulrike Martens\IEEEauthorrefmark{3}, and
Markus M\"{u}nzenberg\IEEEauthorrefmark{3}} %,~\IEEEmembership{Fellow,~IEEE}}
\IEEEauthorblockA{\IEEEauthorrefmark{1}Unit\'{e} Mixte de Physique CNRS/Thales,
Universit\'{e} Paris Sud, 1 Avenue Augustin Fresnel, 91767 Palaiseau, France}
\IEEEauthorblockA{\IEEEauthorrefmark{2}I. Physikalisches Institut, Georg-August-Universit\"{a}t G\"{o}ttingen, Friedrich-Hund-Platz 1, 37077 G\"{o}ttingen, Germany}
\IEEEauthorblockA{\IEEEauthorrefmark{3}Institut f\"{u}r Physik, Ernst-Moritz-Arndt-Universit\"{a}t Greifswald, Felix-Hausdorff-Str. 6, 17489 Greifswald, Germany}
%\IEEEauthorblockA{\IEEEauthorrefmark{4}Tyrell Inc., 123 Replicant Street, Los Angeles, CA 90210 USA}% <-this % stops an unwanted space
\thanks{Manuscript received April 21, 2017; revised April 21, 2017. 
Corresponding author: S. Wittrock (email: steffen.wittrock@u-psud.fr).}}

\markboth{IEEE Transactions on Magnetics, Vol. 53, No. 11, November 2017}%
{Shell \MakeLowercase{\textit{et al.}}: Bare Demo of IEEEtran.cls for IEEE Transactions on Magnetics Journals}



% for Transactions on Magnetics papers, we must declare the abstract and
% index terms PRIOR to the title within the \IEEEtitleabstractindextext
% IEEEtran command as these need to go into the title area created by
\IEEEtitleabstractindextext{%
\begin{abstract}
Spin currents offer a way to control static and dynamic magnetic properties, and therefore they are crucial for next-generation MRAM devices or spin-torque oscillators.
Manipulating the dynamics is especially interesting within the context of photo-magnonics. In typical $3d$ transition metal ferromagnets like CoFeB, the lifetime of light-induced magnetization dynamics is restricted to about 1 ns, which e.g. strongly limits the opportunities to exploit the wave nature in a magnonic crystal filtering device. 
Here, we investigate the potential of spin-currents to increase the spin wave lifetime in a functional bilayer system, consisting of a heavy metal (8 nm of $\beta$-Tantalum (Platinum)) and 5 nm CoFeB. 
Due to the spin Hall effect, the heavy metal layer generates a transverse spin current when a lateral charge current passes through the strip.
Using time-resolved all-optical pump-probe spectroscopy, we investigate how this spin current affects the magnetization dynamics in the adjacent CoFeB layer.
We observed a linear spin current manipulation of the effective Gilbert damping parameter for the Kittel mode from which we were able to determine the system's spin Hall angles. Furthermore, we measured a strong influence of the spin current on a high-frequency mode. We interpret this mode an an exchange dominated higher order spin-wave resonance. Thus we infer a strong dependence of the exchange constant on the spin current.
% on another ocurring, standing spin wave mode, driven by an observed dependence of the exchange constant on the spin current.

%Here, i.a., we investigate the potential of spin-currents to increase the spin wave lifetime in a simple trilayer system, consisting of a heavy metal (8 nm of $\beta$-Tantalum (Platinum)), 5 nm CoFeB, capped by 3 nm Ruthenium. The samples were grown in UHV by magnetron sputtering (Ta, CoFeB) and E-beam evaporation (Pt, Ru), and subsequently patterned into micron-sized conduction strips using E-beam-lithography. Due to the spin Hall effect, the heavy metal layer generates a transverse spin current when a lateral charge current passes through the strip. Using time-resolved all-optical pump-probe spectroscopy, we investigate how this spin current affects the magnetization dynamics in the adjacent CoFeB layer.

\end{abstract}

}

\maketitle

\IEEEdisplaynontitleabstractindextext

% \fi
\IEEEpeerreviewmaketitle

\section{Introduction}

\IEEEPARstart{T}{he} spin-transfer effect describes the transfer of spin angular momentum to a ferromagnet's magnetization from an injected spin polarized current.
Since its prediction by L. Berger \cite{Berger1984}, this effect has experienced a high research interest as it permits to manipulate and control the magnetization of a thin ferromagnetic (FM) layer.
Especially when the resulting spin transfer torque is collinear with the damping torque, magnetic dissipation can be controlled. Thereby the life time of spin wave dynamics can be drastically enhanced.
%Especially its influence on the effective Gilbert damping parameter $\tilde{\alpha}$ of magnetization dynamics due to the collinearity of their torque terms has attracted attention, since the restriction of spin wave lifetimes can be overcome.

Among the possible methods of creating the necessary spin current, the exploitation of the spin Hall effect has become a powerful mean since its first observation only a decade ago \cite{Kato2004,Wunderlich2004,Day2005,Wunderlich2005}. Governed by spin-orbit coupling phenomena, it generates a spin current $j_s$ from a transverse charge current $j_e$ without any need for neither a ferromagnet nor an external magnetic field. The efficiency of the conversion process can be described by the spin Hall angle (SHA) $\Theta_{SH}=j_s/j_e$.

Here, we investigate the photo-induced magnetization dynamics in a few nanometer thin soft magnetic layer consisting of amorphous metallic cobalt iron boron alloy (Co$_{20}$Fe$_{60}$B$_{20}$) under influence of a strong spin current generated by the SHE in an adjacent heavy metal film made of platinum (Pt) or tantalum (Ta). The sputter conditions for Ta were chosen such that the film has grown in the high resistive $\beta$-phase for which a high SHA has recently been reported \cite{Liu2012}.
%The measurements were performed using the powerful tool of time resolved magnetooptical Kerr effect. Compared to more established methods like BLS or ST-FMR, the benefit of our experimental approach lies in the possibility to investigate (sub-)picosecond dynamics \textcolor{red}{which we showed is also being manipulated by spin currents (not part of this manuscript).} 


\section{Experimental}

The samples consist of two functional thin layers (fig. \ref{subfig:experimental_schematic}): $8\,$nm Pt or $\beta$-Ta as a SHE-material generating the spin current and $5\,$nm of ferromagnetic amorphous CoFeB, into which the spin current is being injected in order to manipulate its magnetization dynamics. The layer stack is complemented by $3\,$nm of Ru as a capping layer. CoFeB and Ta are deposited by argon ion sputtering and Pt and Ru by e-beam evaporation. All preparation steps are conducted in situ in ultra-high vacuum.
Subsequent structuring of the samples by e-beam lithography enables the electrical contacting and generation of a high charge current density (fig. \ref{subfig:experimental_pattern}). 
%aims to allow electrical contacting and furthermore generation of a high charge current density $j_e$ (fig. \ref{subfig:experimental_pattern}). 
%Fig. \ref{subfig:experimental_pattern} shows the patterned structure, fig. \ref{subfig:experimental_schematic} the processes in the functional layers: The optically induced magnetization dynamics in manipulated by an injected spin current generated by SHE from a transversal charge current sent through the patterned conduction strip.

%\begin{figure}[!t]
%\centering
%\includegraphics[width=2.5in]{figures/SOT+M-praezession.pdf}
%\caption{Simulation results for the network.}
%\label{fig_sim}
%\end{figure}

%\renewcommand\thesubfigure{(\alph{subfigure})}
%\newcommand\p@subfigure{(\thefigure}

\newsavebox{\leftbox}

\begin{figure}[hbt!]
  \savebox{\leftbox}{
 \noindent \begin{minipage}[b]{0.22\textwidth}
    \centering

    \subcaptionbox{ \label{subfig:experimental_schematic}}{\includegraphics[width=0.78\textwidth]{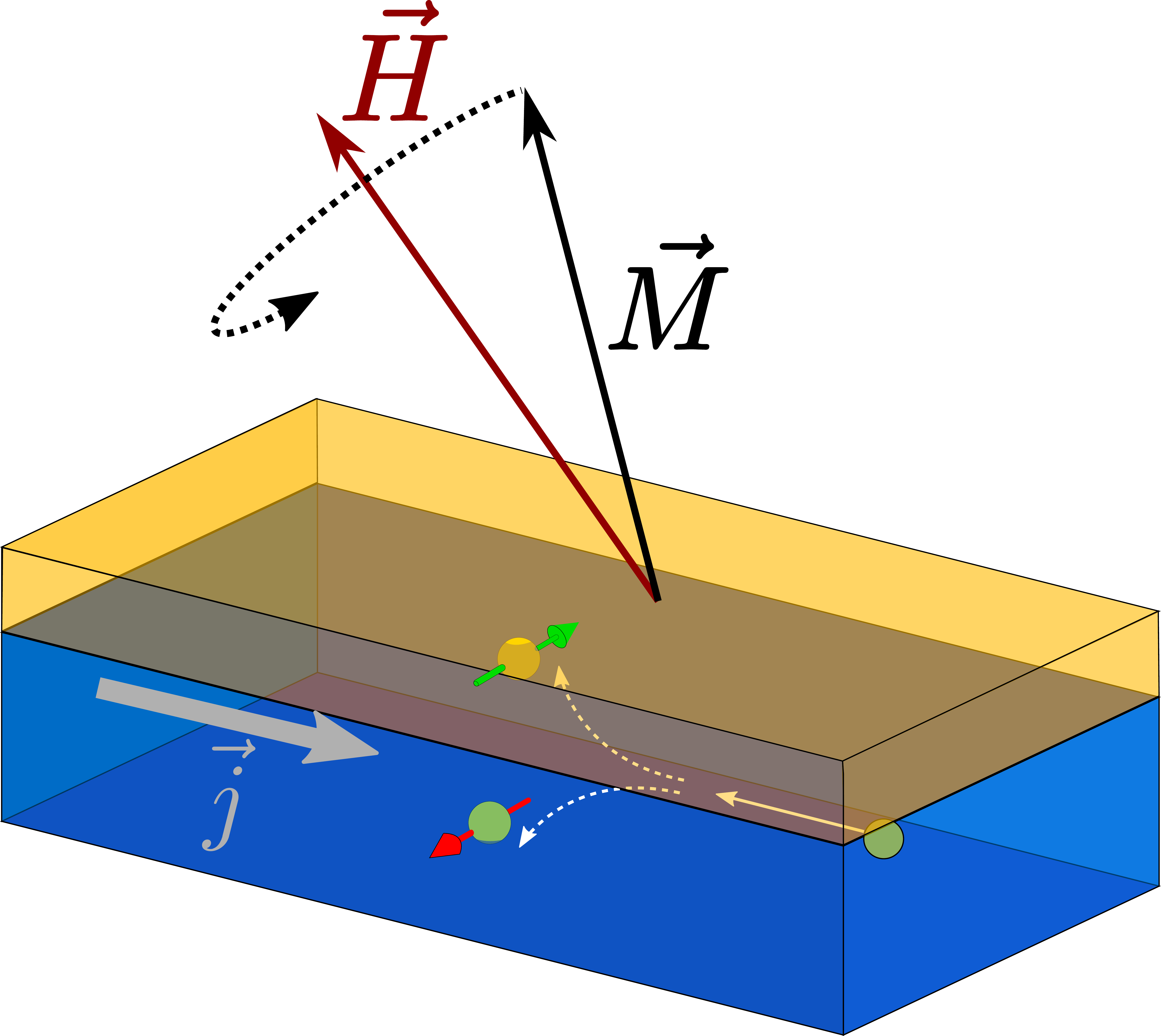} }
    \subcaptionbox{ \label{test} \label{subfig:experimental_pattern}}{\includegraphics[width=0.22\textwidth, angle=90]{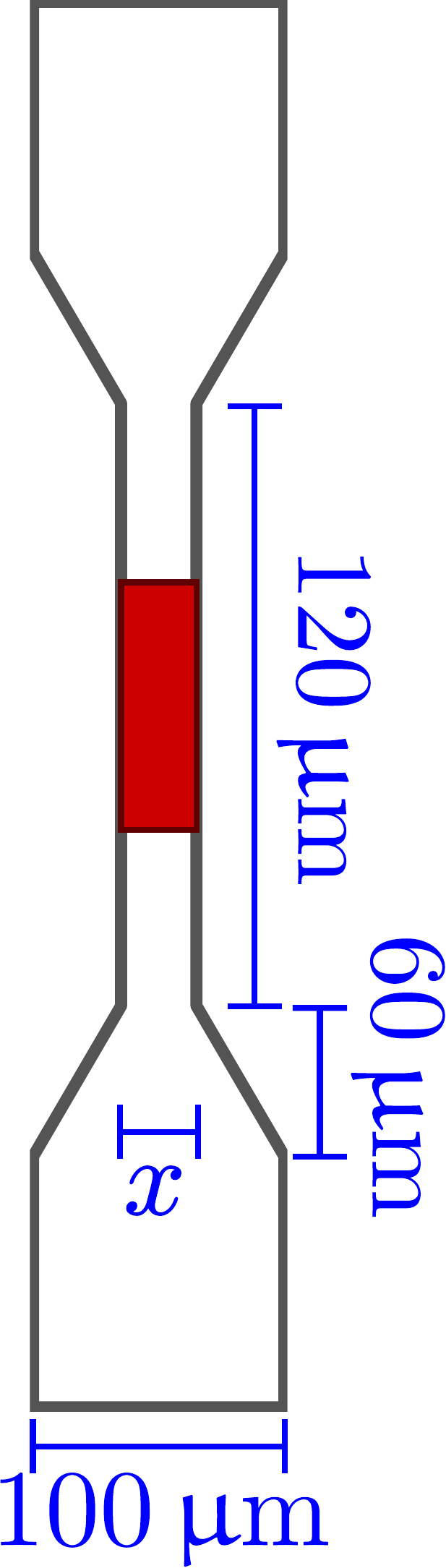} }  
  \end{minipage}%
  } \usebox{\leftbox}
 \noindent \begin{minipage}[b][\ht\leftbox][b]{0.28\textwidth}
    \centering		%addtocounter{subfigure}{+2}  setcounter{subfigure}{2}
    \subcaptionbox{ \label{subfig:experimental_setup}}{\includegraphics[width=\textwidth]{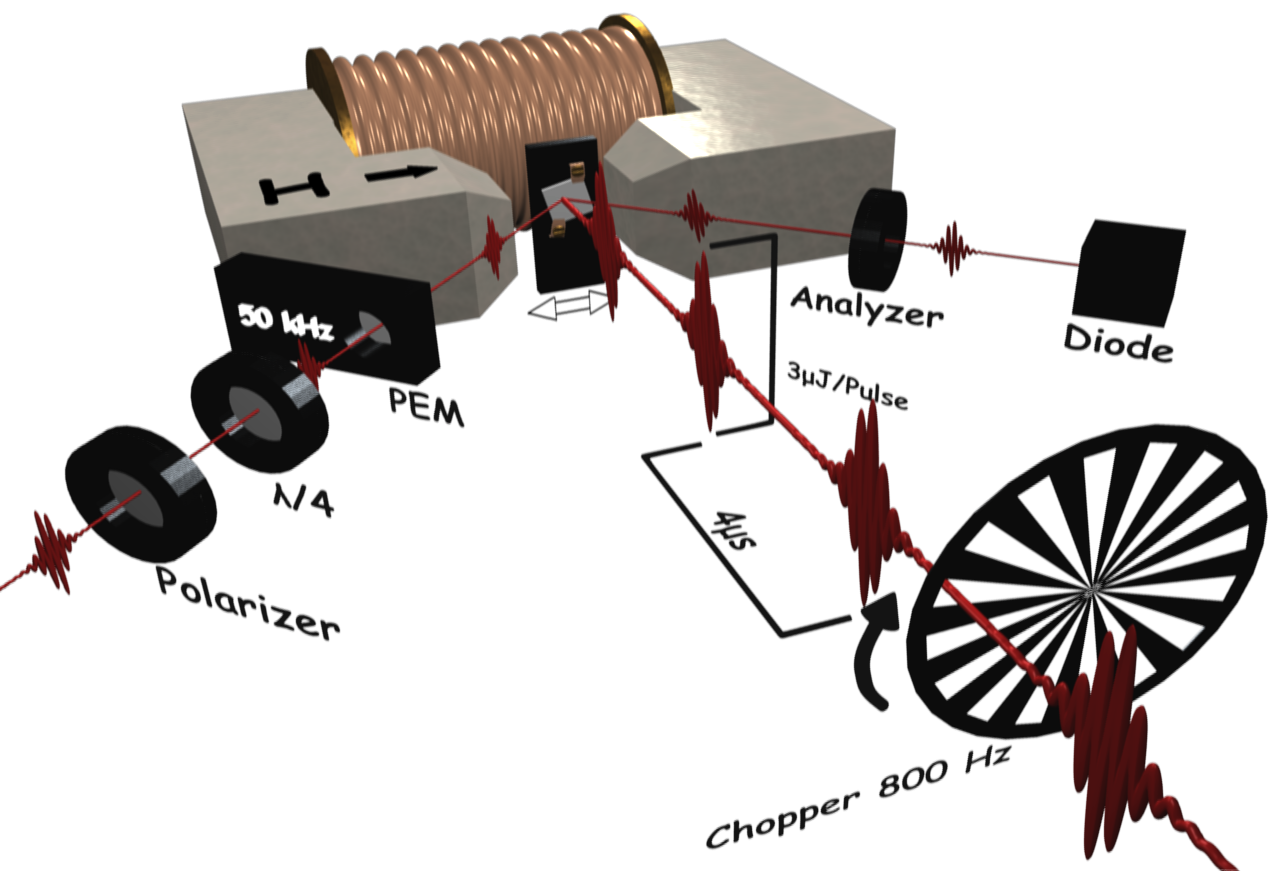} }
  \end{minipage}
  \caption{Experimental characteristics. (a) Schematic processes in the functional layer stack of a SHE material (blue) and the ferromagnetic CoFeB (yellow). (b) Patterned sample structure, the width $x$ was $12\,\upmu$m for the results shown here. The red marked area was excited by the laser spot. (c) Pump-probe setup with ratio of powers $P_{pump}:P_{probe}=95:5$, time resolution is realized by a delay stage, a double modulation technique of photoelastic modulator (PEM) and chopper frequency was used \cite{Koopmans2003}.}
  \label{fig:experimental}
\end{figure}

We used time resolved pump-probe spectroscopy exploiting the magnetooptical Kerr effect to excite and measure magnetization dynamics (schematical setup in \ref{subfig:experimental_setup}). The laser pulses with central wave length $\lambda=800\,$nm have an autocorrelation length of $\Delta\tau\approx80\,$fs and a repetition rate of $250\,$kHz. The pump spot size was $\approx 60\,\upmu$m providing an optical fluence of $F=15\,\unitfrac{mJ}{cm^2}$. 

The incoming pump pulse induces the dynamics at $\tau=0$ by firstly generating hot electrons, which thermalize on a timescale of $\tau\sim 100\,$fs due to electron-electron-scattering. Further scattering events with phonons and spins lead to energy transfer into the phonon and spin system giving rise to ultrafast demagnetization \cite{Beaurepaire1996}. Caused by the high change in temperature on the time scale of $\sim 1\,$ps, the local anisotropy constant of the ferromagnetic material and thus the effective magnetic field $\vec{H}_{eff}$ is changed. With $\vec{H}_{eff}$ rereaching its equilibrium position after a few picoseconds, the magnetization dynamics corresponding to the Landau-Lifshitz-Gilbert-equation (LLG) is excited. 
In order to subsequently trigger a coherent precession of the magnetization, an external field ($145\,$mT) was applied at an out-of-plane angle of $\varphi=35\,^{\circ}$, transversal to $\vec{j}_e$. 

%With $\vec{H}_{eff}$ rereaching its equilibrium position after a few picoseconds, the magnetization $\vec{M}$ starts precessing around the effective field corresponding to the Landau-Lifshitz-Gilbert-equation (LLG). 

%\begin{figure}[!tbh]
%\centering
%
%\setlength{\fboxsep}{0pt}
%\setlength{\fboxrule}{.1pt}
%
%%\fbox{
%%\noindent\begin{minipage}[t]{0.5\textwidth}
%
%
%\newlength{\boxheight}
%
%
%%\fbox{
%\begin{adjustbox}{minipage=[t]{0.22\textwidth},gstore totalheight=\boxheight , margin=0}
%%\begin{minipage}[t][][b]{0.22\textwidth}
% \vspace{0pt}
%\subfloat[]{  \begin{varwidth}{\linewidth}
%	  \includegraphics[width=0.2\textwidth , angle=90]{figures/Ta-Bahn_2.pdf}\\ %0.1
%      \includegraphics[width=0.7\textwidth]{figures/SOT+M-praezession.pdf}	%0.3
%      \end{varwidth} 
%\label{fig:bla}}
%%
%\end{adjustbox} 
%%}
%%\usebox{\LeftBox}
%%
%%\fbox{
%\begin{adjustbox}{minipage=[t][\boxheight]{0.22\textwidth}}
% \vspace{0pt}
%\subfloat[]{\includegraphics[width=\textwidth]{figures/TRMOKE-setup.png}%
%\label{fig:blubb}}
%
%\end{adjustbox}
%%}
%%\end{minipage}
%% }
%
%\caption{Simulation results for the network.}
%\label{fig:sim}
%\end{figure}

\section{Results - nanosecond timescale}

\begin{figure}[hbt!]
 \noindent \begin{minipage}[b]{0.49\textwidth}
    \centering
    \subcaptionbox{\label{subfig:rohdaten}}{\includegraphics[width=0.7\textwidth]{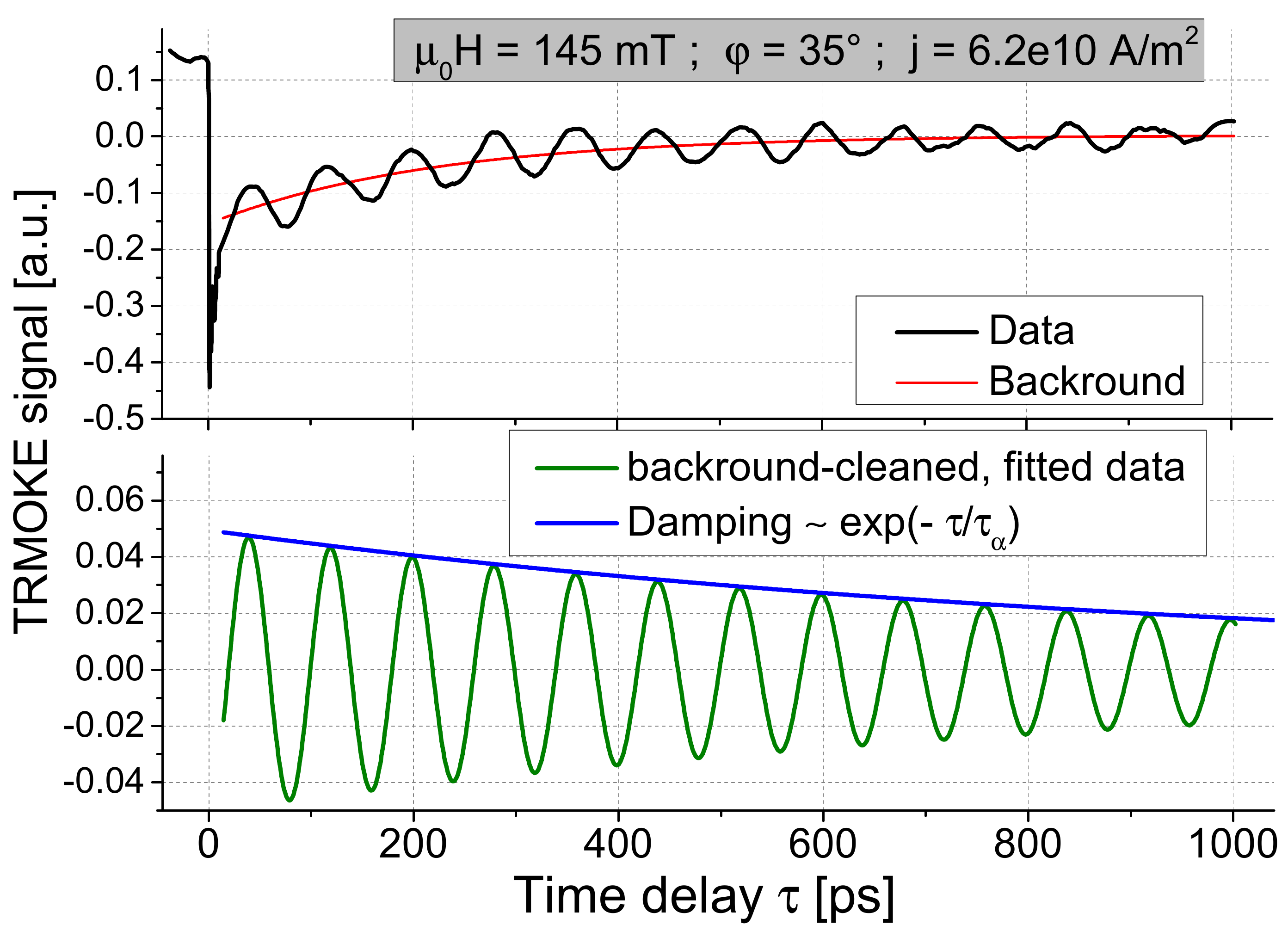}} 
  \end{minipage}% 
  \newline
 \noindent \begin{minipage}[b][][b]{0.5\textwidth}
    \centering
    \subcaptionbox{\label{subfig:f_j}}{\includegraphics[width=0.48\textwidth]{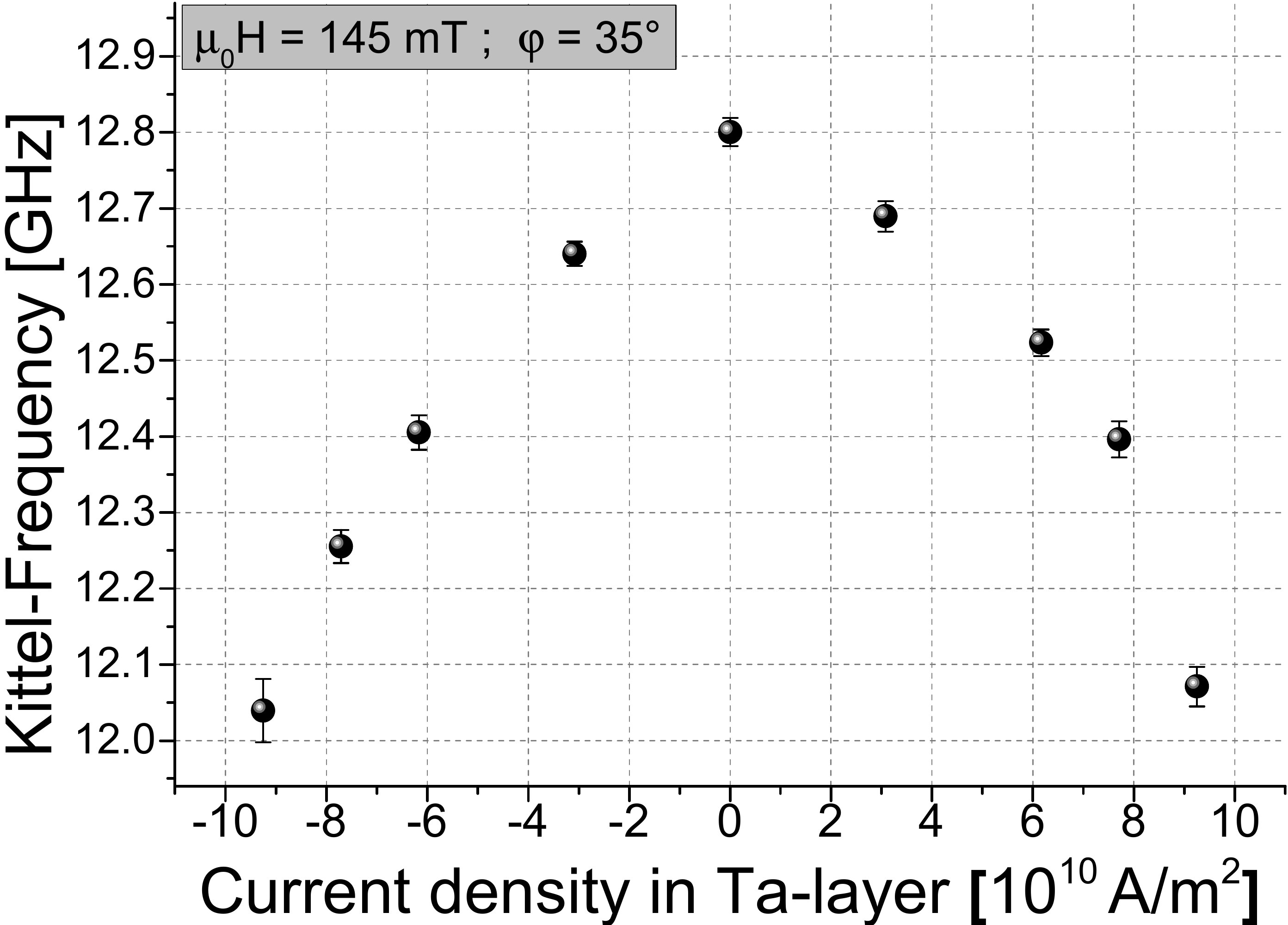} }
    \subcaptionbox{\label{subfig:M_j}}{\includegraphics[width=0.48\textwidth]{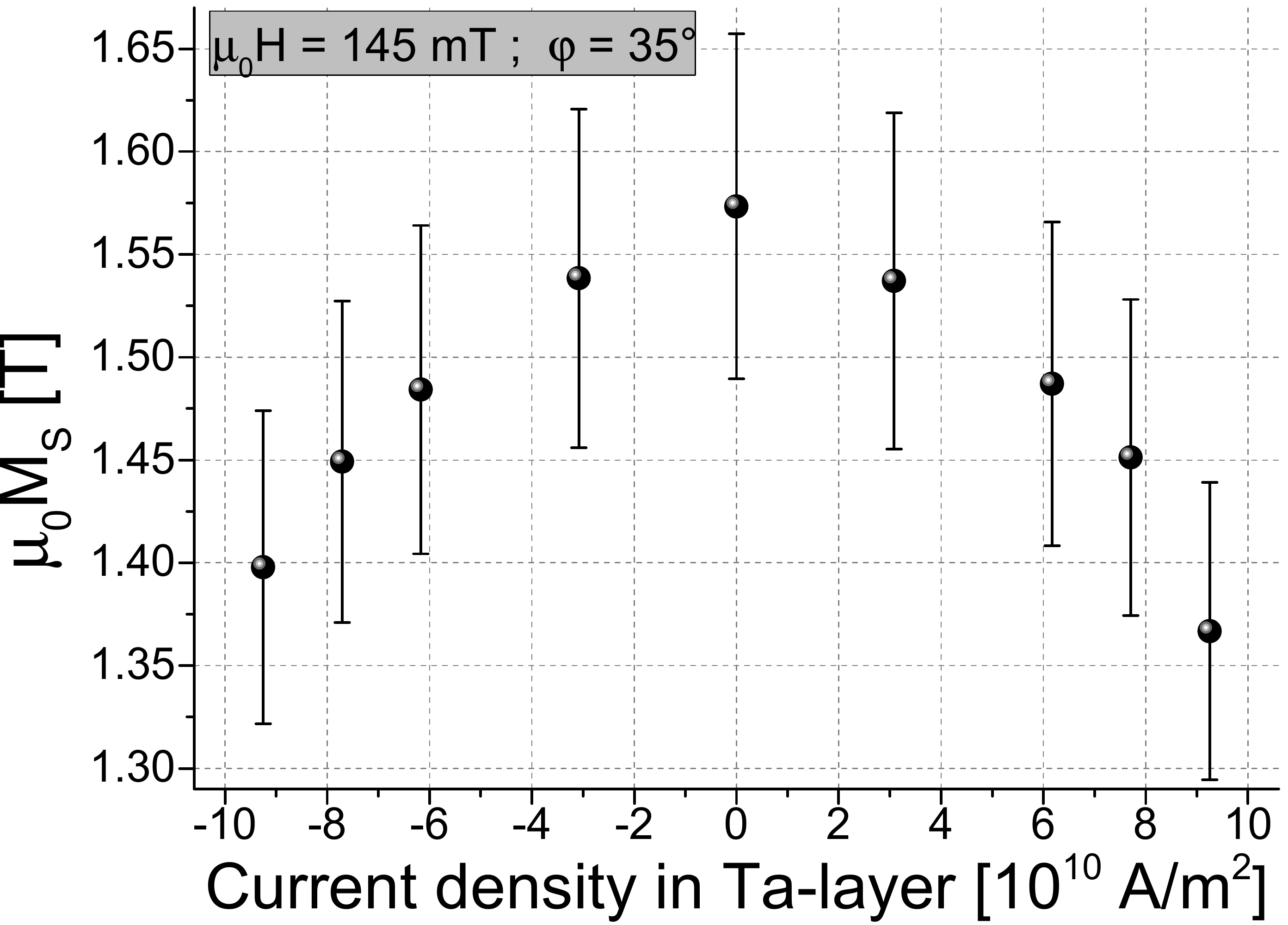} }
  \end{minipage}
  \caption{(a) Typical measurement data and Kittel mode after substraction of exponential background. The TRMOKE-signal is proportional to the magnetization. Dependence of (b) Kittel frequency and (c) saturation magnetization on the current density in the Ta-layer. }
  \label{fig:results1}
\end{figure}

Besides the dominating coherent, in spatially homogeneous geometries called Kittel mode oscillations, also incoherent phonons and magnons contribute to the signal and give rise to a certain background \cite{Djordjevic2006, Beaurepaire1996}, which can be modelled by exponential functions.
%After the pump pulse induced an anisotropy pulse of a few picoseconds and the sample was demagnetized, at first incoherent magnons show up until their damping makes the Kittel mode dominate the relaxation process. It is remarked that, next to the dominating magnetic, also phononic processes contribute to the signal \cite{Djordjevic2006}. 
%The background can be modelled by exponential functions. 
Substracting it from the raw data (fig. \ref{fig:results1}) highlights the magnetic oscillations.
These are analysed with respect to frequency and damping for different current densities passing through the SHE-material.

\subsection{Kittel frequencies, magnetization, and Oersted field}

The dependence of the Kittel frequency on $j$ is shown in fig. \ref{subfig:f_j}. The mainly parabolic behaviour can be attributed to the reduction of the saturation magnetization due to Joule-heating. The observable asymmetries for opposite current directions can be related to the Oersted field produced by the current, or to the presence of a field-like torque. Analysing the asymmetries, an Oersted field of $H_{Oe}=aj$ with $a=(1.23 \pm 0.04)\cdot 10^{-8}\,$m is determined which agrees well with theoretical estimations. Even if present, the field-like torque must be much smaller than the effect arising from the Oersted field.
Taking also into account the in-plane applied magnetic field component $H_x$, the saturation magnetization can be determined from the Kittel formula $\omega = \gamma \mu_0 \sqrt{H_x (H_x +M_S)}$ (fig. \ref{subfig:M_j}). Note that for the given field geometry, the magnetization's out-of-plane component is only around $2\,$\%, as is estimated by micromagnetic simulations; therefore this approximation is valid. Besides Joule-heating, also the energy deposition of the pump pulse heats up the sample locally to around $400$-$450\,$K at a time scale of up to $1\,$ns.
Thus the saturation magnetization at $j=0\,$A/m$^2$ is slightly lower than the room temperature value of $\mu_0 M_S=1.63\,$T, determined by a vibrating sample magnetometer \cite{Mansurova2016}. 
The Joule-heating effect leads to a temperature increase especially in Ta of up to $200\,$K at a maximum current density of $j_{Ta}=9.3\cdot 10^{10}\,$A/m$^2$.

\subsection{Effective Gilbert damping parameter of the Kittel mode}

From the exponential decay time $\tau_{\alpha}$ of the Kittel mode (fig. \ref{subfig:rohdaten}) and taking into account the full in-plane magnetic field $H_x=H_{ext}\cos(\varphi)+H_{Oe}$, and the current dependence of the magnetization, we calculate the effective Gilbert damping parameter using:
% the determined saturation magnetization, and the in-plane magnetic field $H_x=H_{ext}\cos(\varphi)+H_{Oe}$, the effective Gilbert damping parameter can be calculated:
\begin{align}
\alpha_{Kittel}= \left( \tau_{\alpha} \gamma\mu_0 \left( H_x + \frac{M_S}{2} \right) \right)^{-1} ~~~.
\end{align}

\begin{figure}[hbt!]
 \noindent \begin{minipage}[b]{0.245\textwidth}
    \centering
    \subcaptionbox{\label{subfig:alpha_Ta}}{\includegraphics[width=\textwidth]{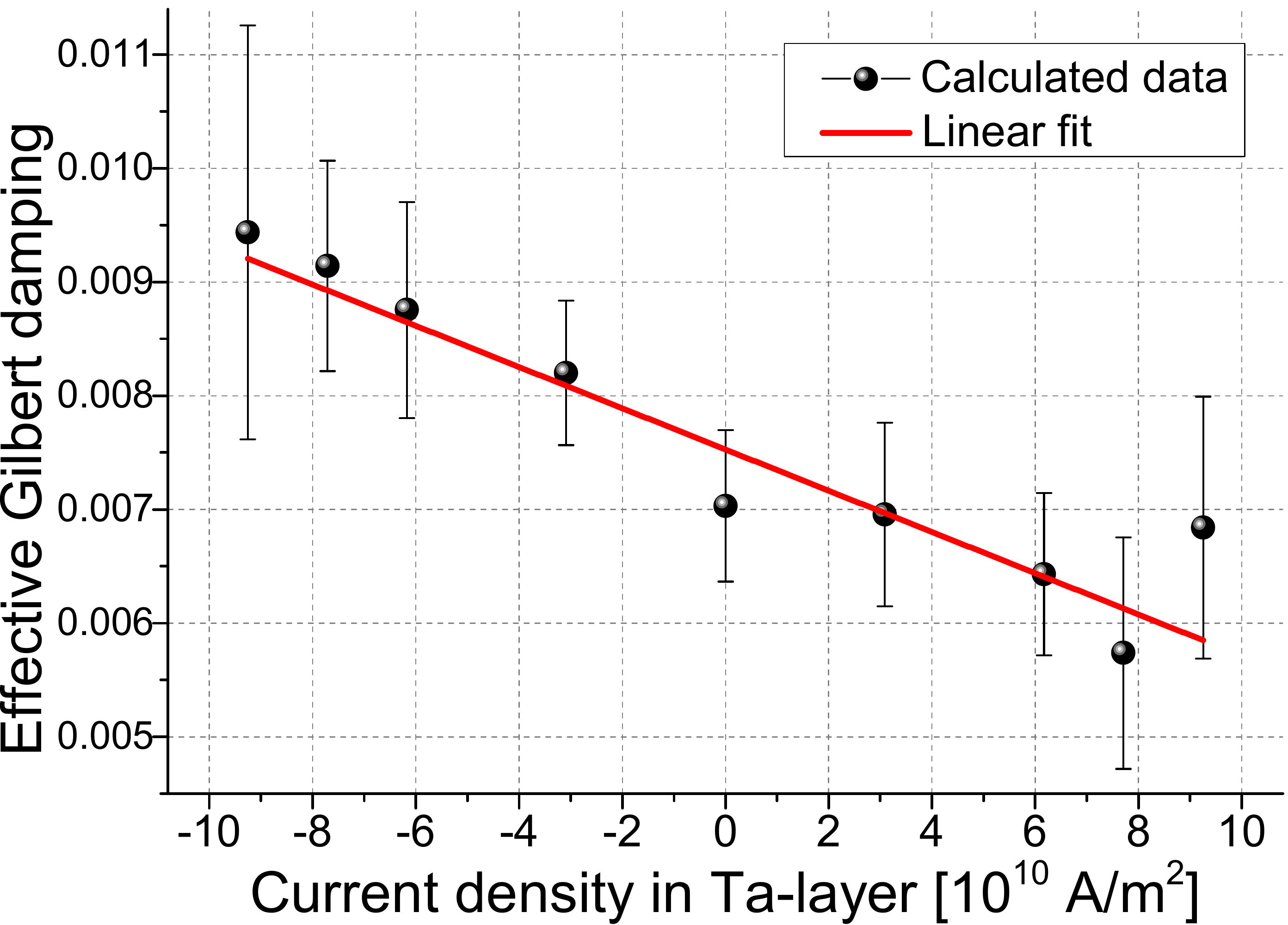}} 
  \end{minipage}% 
 \noindent \begin{minipage}[b][][b]{0.245\textwidth}
    \centering
    \subcaptionbox{\label{subfig:alpha_Pt}}{\includegraphics[width=\textwidth]{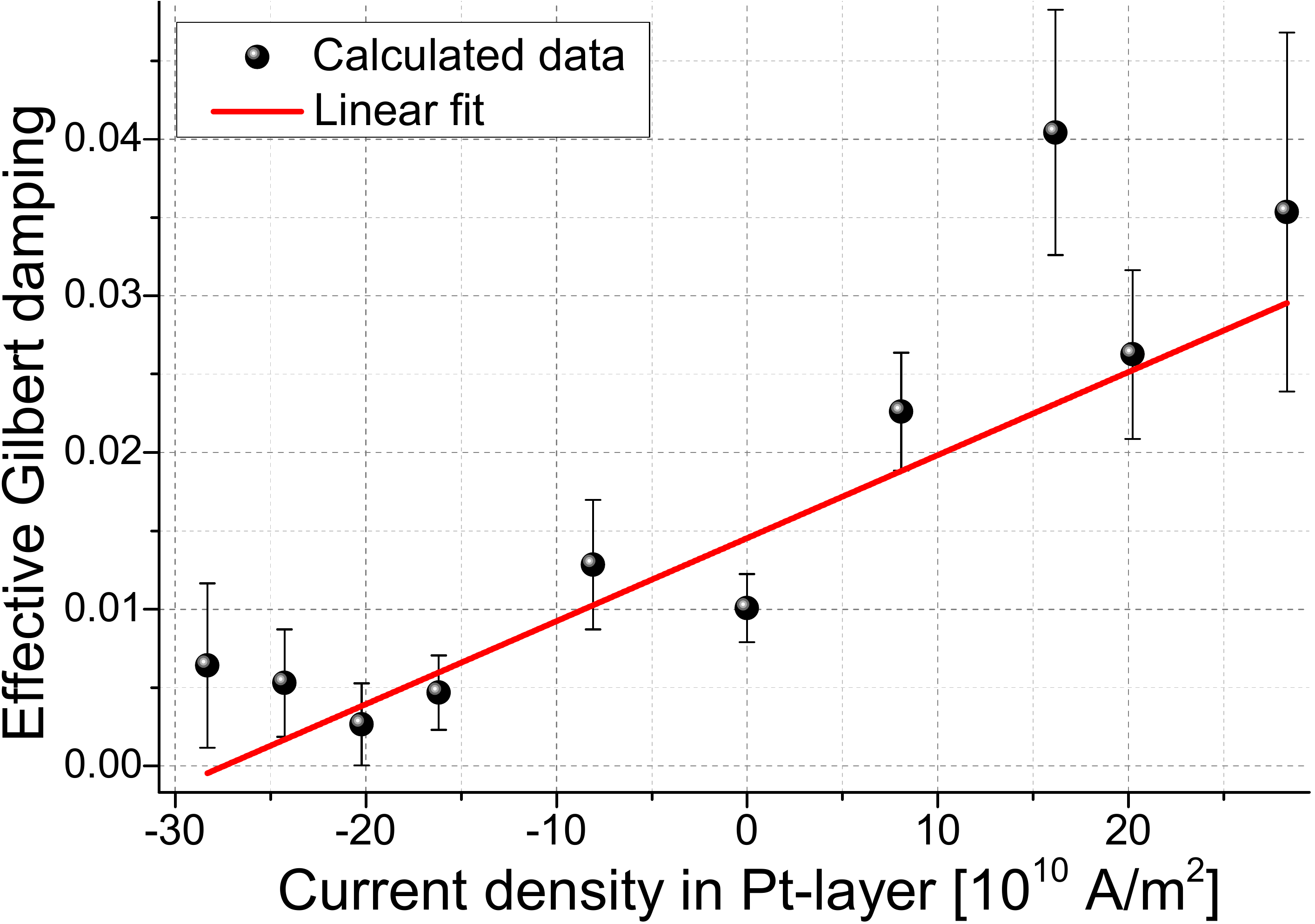} }
  \end{minipage}
  \caption{Current dependence of the effective Gilbert damping parameter for (a) the Ta based sample and (b) the Pt based sample. From the linear fit, the SHE efficiency can be extracted.  }
  \label{fig:results2}
\end{figure}

Results for the two SHE-materials Ta and Pt are shown in fig. \ref{fig:results2}. 
We determined the SHA $\Theta_{SH}$ by fitting the formula:
%They confirm the different sign in the spin-Hall-angle (SHA) $\Theta_{SH}$, determined by a linear fit through:
\begin{align}
\tilde{\alpha} = \alpha + j\cdot \Theta_{SH} \cdot \frac{\hbar}{2ed\mu_0 M_S H_{x}} ~~~. 
\end{align}
Note that this expression was derived from linearizing the Landau-Lifshitz-Gilbert-Slonczewski-equation (LLGS).

Averaging a few datasets gives the SHAs of:
\begin{align*}
\Theta_{SH}^{Ta}&=-0.043 \pm 0.011 ~~~, \intertext{and} \Theta_{SH}^{Pt}&=0.086\pm 0.012 ~~~.
\end{align*}
These results lie within the values obtained by other groups \cite{Sinova2015} and in particular confirm the different sign to be expected for Pt \cite{Liu2011, Ando2008} and Ta \cite{Morota2011, Liu2012}.

\subsection{Spin-pumping effect}

The spin mixing conductance $g^{\uparrow\downarrow}$ of a layer combination can be estimated through comparison of the effective damping parameters $\tilde{\alpha}$ of two layer stacks involving the same ferromagnetic material. Using Ta and Pt as SHE-materials and CoFeB as the ferromagnet, the spin mixing conductance of Ta can be calculated using:
\begin{align}
g_{Ta}^{\uparrow\downarrow} = - \frac{4\pi M_S d\cdot \Delta \tilde{\alpha}}{\hbar \gamma} + g_{Pt}^{\uparrow\downarrow}
\end{align}
$\Delta\tilde{\alpha}=\tilde{\alpha}_{Pt}-\tilde{\alpha}_{Ta}$ denotes the difference of the effective magnetic damping constants of the $d=5\,$nm thick CoFeB layer for the two different adjacent SHE-materials. Average values of a few measurements give $\tilde{\alpha}_{Pt}=(1.20\pm 0.15)\cdot 10^{-2}$ and $\tilde{\alpha}_{Ta}=(0.69 \pm 0.05)\cdot 10^{-2}$ for $j=0$. The Pt/CoFeB  spin mixing conductance of $g_{Pt}^{\uparrow\downarrow}=(4\pm 1)\cdot 10^{19}\,$m$^{-2}$ \cite{Calaforra2015} was used as a reference value.

We evaluate the spin mixing conductance of Ta/CoFeB to $g_{Ta}^{\uparrow\downarrow}=(1.9\pm 1.2)\cdot 10^{19}\,$m$^{-2}$.
The smaller value for Ta/CoFeB compared to Pt/CoFeB is expected as similar values were already obtained in Ta/NiFe \cite{Montoya2016} resp. Pt/NiFe \cite{Czeschka2011} bilayers. 
In conclusion, this analysis shows that TRMOKE is also a suitable method to determine the spin mixing conductance which is an important parameter for heavy metal/FM based bilayer systems.

%This value is conclusive as a high spin mixing conductance is especially found in systems including heavy metals with low spin flip relaxation time $\tau_{sf}$ which can be roughly estimated as $\tau_{sf}\sim 1/Z^4$.

%\IEEEpubidadjcol

\section{Results - picosecond timescale}

\begin{figure}[hbt!]
 \noindent \begin{minipage}[b]{0.5\textwidth}
    \centering
    \subcaptionbox{\label{subfig:TRMOKE-kurz}}{\includegraphics[width=0.56\textwidth]{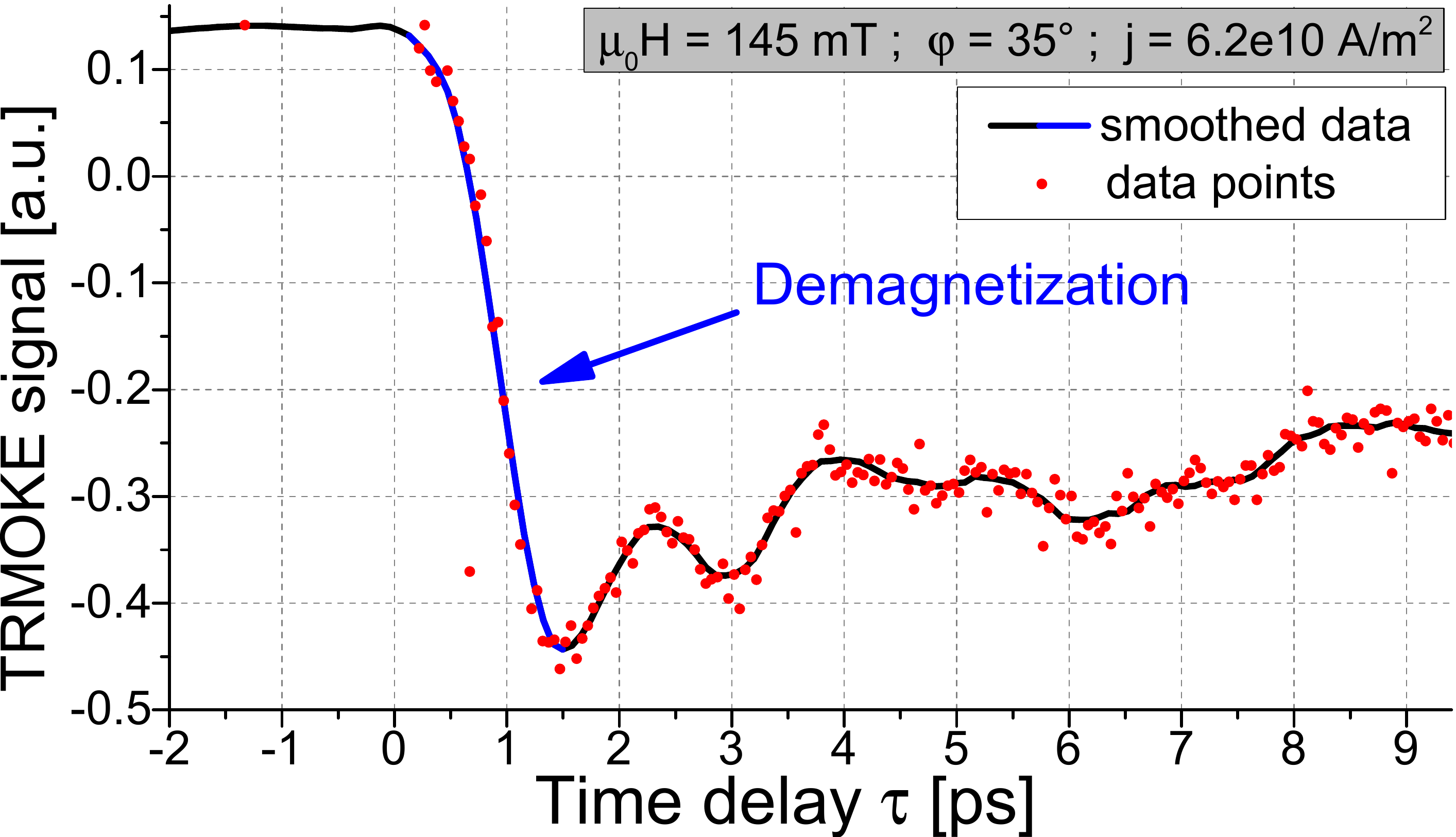}} 
  \end{minipage}% 
  \newline
  \noindent \begin{minipage}[b][][b]{0.24\textwidth}
    \centering
    \subcaptionbox{\label{subfig:PSSW-Fit}}{\includegraphics[width=\textwidth]{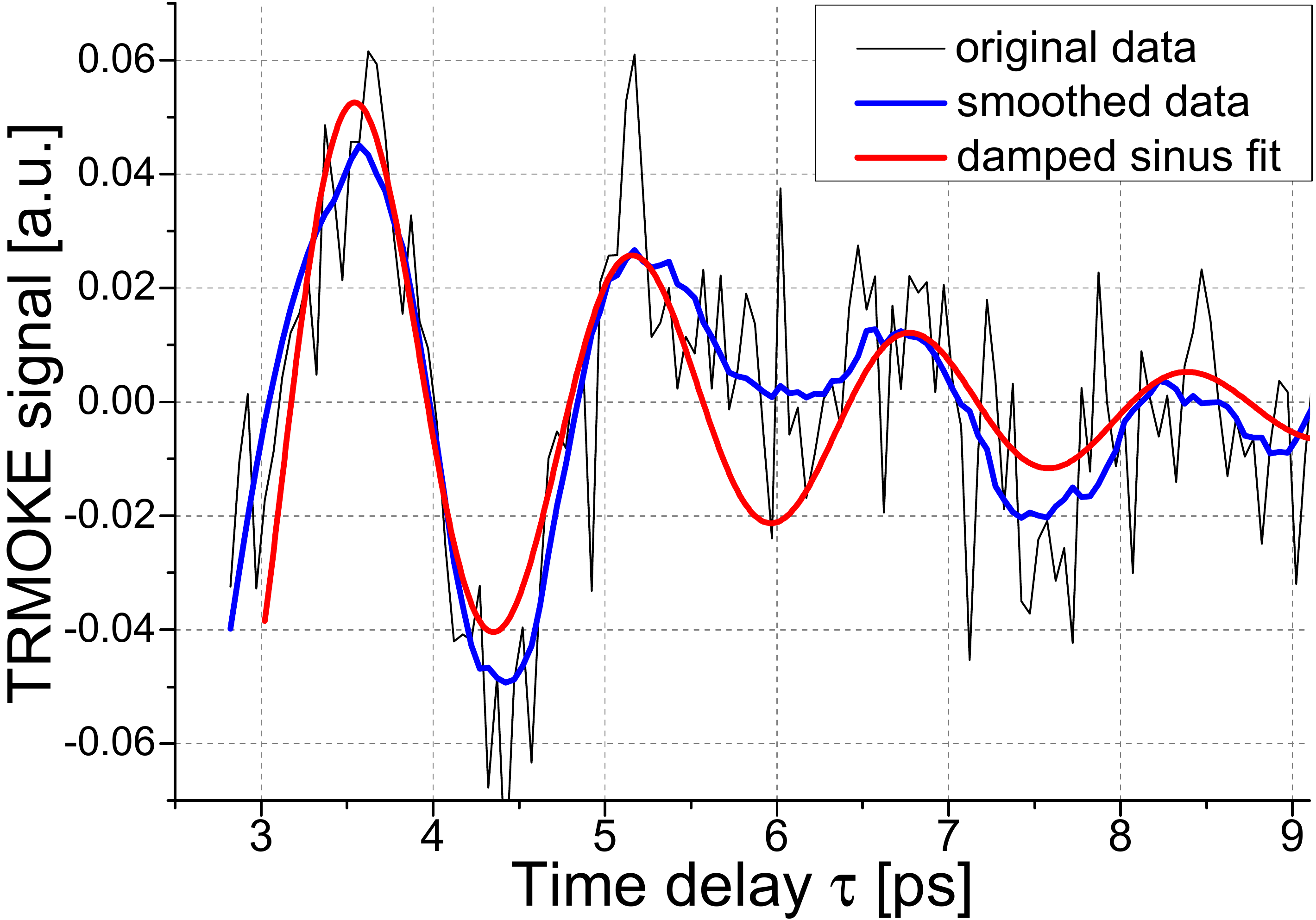} }
  \end{minipage}
 \noindent \begin{minipage}[b][][b]{0.24\textwidth}
    \centering
    \subcaptionbox{\label{subfig:frequ_PSSW}}{\includegraphics[width=\textwidth]{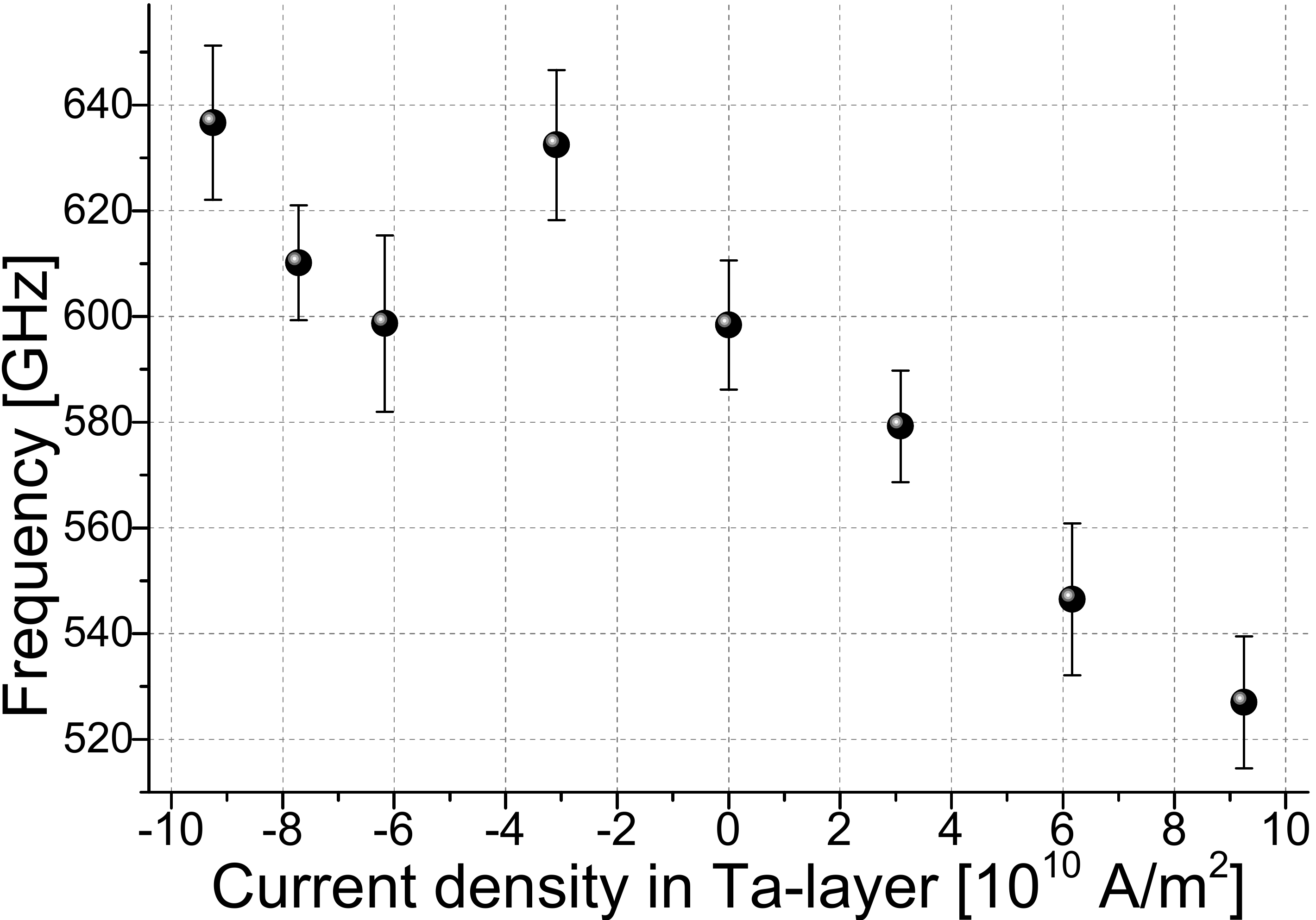} }
  \end{minipage}
  \newline
  \noindent \begin{minipage}[b]{0.245\textwidth}
    \centering
    \subcaptionbox{\label{subfig:austauschsteifigkeit}}{\includegraphics[width=\textwidth]{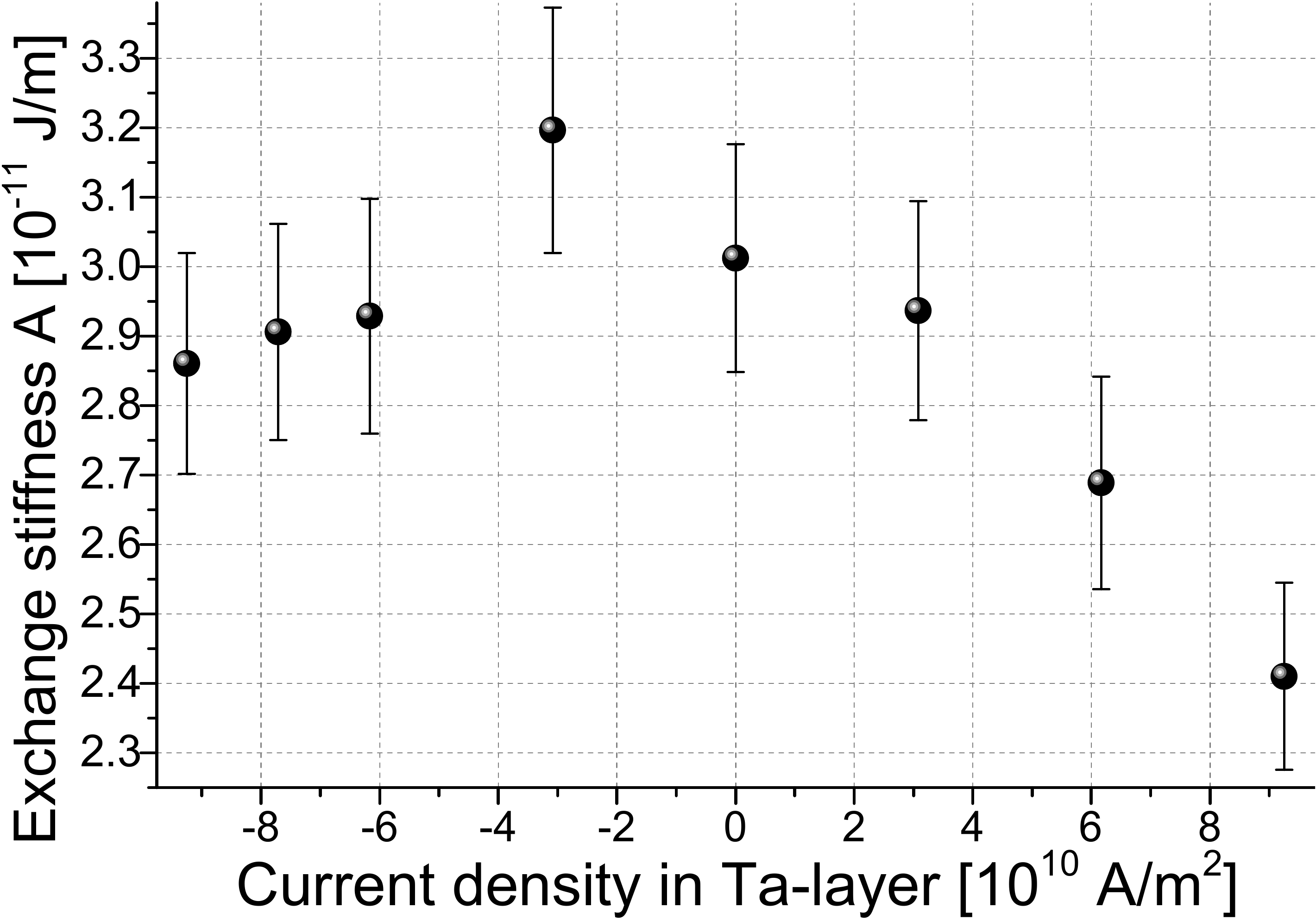}} 
  \end{minipage}% 
 \noindent \begin{minipage}[b][][b]{0.245\textwidth}
    \centering
    \subcaptionbox{\label{subfig:alpha_PSSW}}{\includegraphics[width=\textwidth]{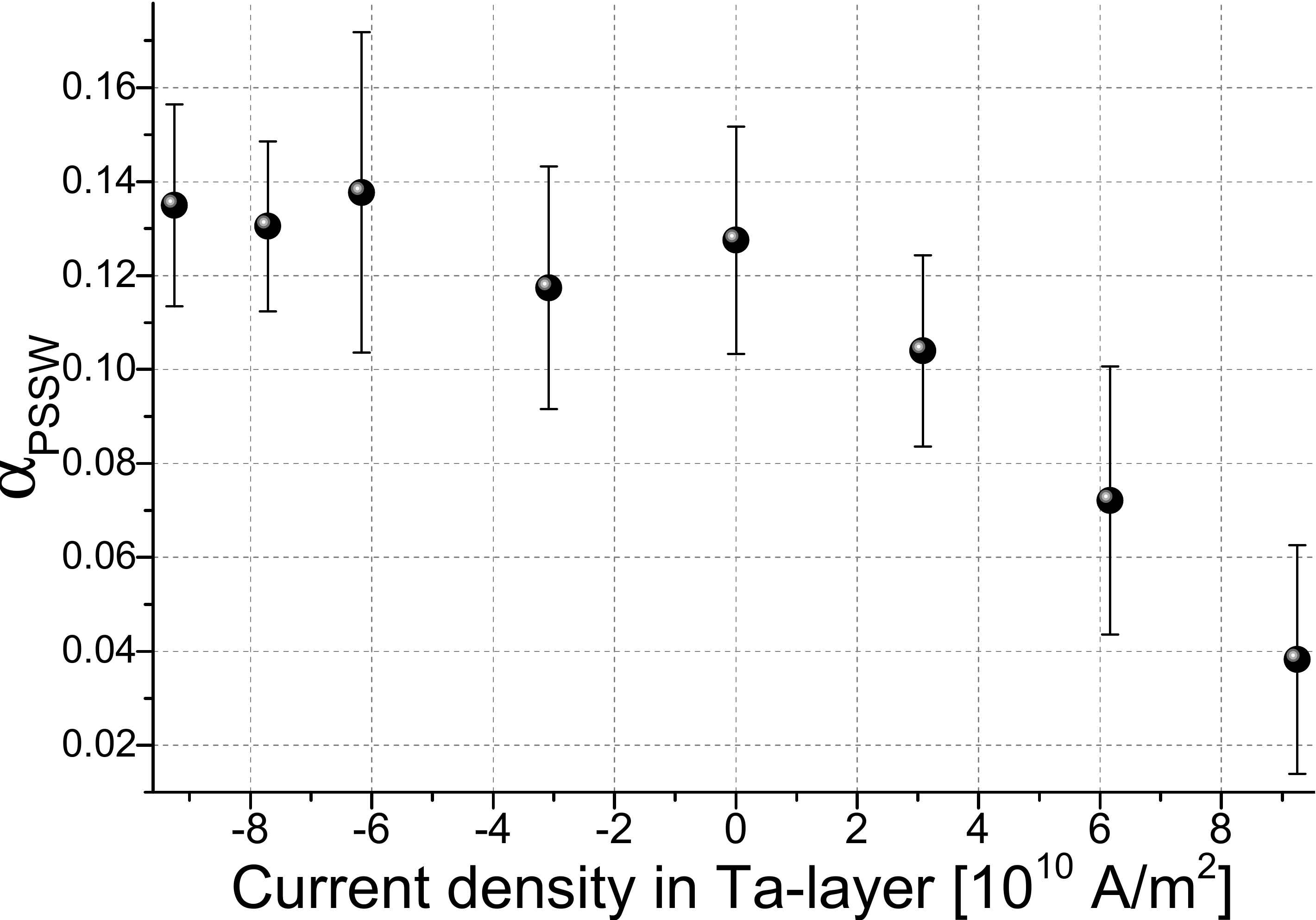} }
  \end{minipage}
  \caption{(a) The first ten picoseconds of the excited dynamics including an ultrafast, highly damped oscillation during remagnetization (b). (c-e) Current dependence of certain derived parameters.}
  \label{fig:results3}
\end{figure}

On the timescale of a few picoseconds, when the magnetization starts relaxating again, we observed a strongly damped, ultrafast oscillation in the terahertz regime (fig. \ref{subfig:frequ_PSSW}), usually existent for one or two periods (fig. \ref{fig:results3}). We identified this oscillation as the well-known perpendicular standing spin-wave (PSSW) mode of first order $n=1$. Analysing its frequencies (fig. \ref{subfig:frequ_PSSW}) and taking into account the saturation magnetization gathered on the nanosecond timescale, especially the exchange stiffness $A$ (fig. \ref{subfig:austauschsteifigkeit}) can be obtained from:
\begin{align}
\omega= \gamma \mu_0 \sqrt{\left(H_x + \frac{2A}{\mu_0 M_S} k^2\right) \left(H_x + \frac{2A}{\mu_0 M_S} k^2 + M_S\right)}
\end{align}
with $k^2=k_z^2= \left(\nicefrac{n\pi}{d}\right)^2$, $n\in \mathbb{N}$ the quantized wavevector normal to the plane.
The exchange stiffness is shown in fig. \ref{subfig:austauschsteifigkeit} and found to depend on the spin current.

Note that pinning at interfaces can alter the effective wave length of the exchange mode. Without further experimental evidence, assuming zero pinning is the simplest model. Since the exchange constant fits to known values determined from TRMOKE experiments on thicker films \cite{Ulrichs2010}, we stick to this model.

The Gilbert damping parameter of the PSSW mode (fig. \ref{subfig:alpha_PSSW}) can again be determined from the oscillation's exponential decay time $\tau_{\alpha}$:
\begin{align}
\alpha_{PSSW}= \left( \tau_{\alpha} \gamma\mu_0 \left(H_x + \frac{2Ak^2}{\mu_0 M_S} + \frac{M_S}{2} \right) \right)^{-1} ~~~.
\end{align}
Especially the exchange term $\sim A k^2/M_S$ proves to be dominant due to the small CoFeB thickness. 
 We attribute the (at least for positive $j_{Ta}$) linear behaviour of $\alpha_{PSSW}$ to the SHE. The curve's slope (fig. \ref{subfig:alpha_PSSW}, $j_{Ta}\geq 0$) possesses the same sign as the Kittel mode damping $\alpha_{Kittel}$. Furthermore, $\alpha_{PSSW}$ is found to be one order of magnitude higher than $\alpha_{Kittel}$. The relative change from $\alpha_{PSSW}\approx 0.04$ for a high positive current density up to $\alpha_{PSSW}\approx 0.13$ for $j=0$ shows a strong dependence on the spin current which is around three times higher than for the Kittel mode.

%Nevertheless, as fig. \ref{subfig:PSSW-Fit} shows, a proper evaluation, especially for high damping and frequencies (so negative current densities), was hard to perform.

\section{Discussion}

According to our experiments, the SHA for a Pt-based layer structure is almost double the SHA of the Ta-system. 
A strong Joule-heating effect occurred especially in the Ta based structure because of its high resistivity. The latter also contributes to an eventual systematic reduction of the tantalum's SHA by up to $10\,$\% assuming the capped Ru displays a small negative SHA of $\Theta_{SH}\approx -0.001$ \cite{Kampfrath2013}. This would lead to a spin current with opposing sign flowing into the CoFeB from above. % opposing spin current in the CoFeB.
% Also due to its high resistivity, it is possible that the SHA of the Ta based structure is systematically reduced by up to $10\,$\% assuming the capped Ru also having a small SHA of $\Theta_{SH}\approx -0.001$ \cite{Kampfrath2013} leading to a reduced spin current in the CoFeB.

The strong spin pumping mechanism and thus increase of the magnetic damping constant especially in the Pt sample is theoretically expected due to platinum's high spin flip probability. Adding an interlayer with high spin diffusion length (such as Cu \cite{Mizukami2002, Sun2013, Sanchez2014}) could solve this issue.

The interesting discovery of the fast, highly damped magnetic oscillation within the first $10\,$ps of the remagnetization process is identified as the PSSW mode of first order. It shows a strong dependence on the injected spin current which influences its frequency, the exchange stiffness and the mode's damping. Note that usually, the PSSW mode is optically excitable and observable in samples, whose layer thickness is higher than the optical penetration depth which is $\sim 30\,$nm in our case.
%Usually, the PSSW mode is observed in materials significantly thicker than the optical penetration depth which is $\sim 30\,$nm in our case.
To obtain clear evidence of this mode's properties, further investigations with enhanced signal to noise ratio are necessary.

\section{Conclusion and Outlook}

In this article we discussed the photoinduced magnetization dynamics under influence of an injected spin current, generated by the SHE.
The powerful tool of time resolved magnetooptical Kerr effect, compared to more established methods like BLS or ST-FMR, allowed to investigate nanosecond as well as (sub-)picosecond dynamics and to get insights into high-frequency and non-equilibrium dynamics so far not in the focus within this context.  
%The measurements were performed using the powerful tool of time resolved magnetooptical Kerr effect. Compared to more established methods like BLS or ST-FMR, the benefit of our experimental approach lies in the possibility to investigate (sub-)picosecond dynamics \textcolor{red}{which we showed is also being manipulated by spin currents (not part of this manuscript).} 

We discussed the influence of Joule-heating on our samples and described the linear manipulation of the Kittel mode's Gilbert damping through a spin current. The spin Hall angles which could be determined for the two different, commonly used SHE materials Ta and Pt, lie within other reported values \cite{Sinova2015}.

We reported a magnetic oscillation at the picosecond timescale which is found to be highly dependent on the spin current.
The exact behaviour of this mode has to be further investigated in future experiments with enhanced signal to noise ratio.
Furthermore, the interesting timescales of the demagnetization process will be adressed in future publications.

Standard methods such as ST-FMR do not allow to address such high-frequency dynamics easily. Our approach enables us to enter this temporal regime, which provides new insights on the action of spin torques on picosecond time scales.
%In addition, recent development in photo-magnonics deals with generation of ultrafast spin currents [paper von Stefan Mathias und co]. In the view of our results, new experimental direction can be envisaged, which elucidate the interplay of such ultrafast currents with additional static currents.

% or


%%%%%%%%%%%%%%%%%%%%%%%%%%%%%%%%%%%%%%%%%\appendices


\section*{Acknowledgment}

The authors acknowledge financial support by the DFG, within the CRC 1073 'Atomic scale control of energy conversion'.

\ifCLASSOPTIONcaptionsoff
  \newpage
\fi

\end{document}